\documentstyle[aps,preprint]{revtex}
\tightenlines

\def \be {\begin{equation}}
\def \eq {\end{equation}}
\def \bee {\begin{eqnarray}}
\def \eqq {\end{eqnarray}}

\def \bea {\begin{array}{c}}
\def \eqa {\end{array}}

\begin{document}
\title{Casimir effect of Graviton and  the Entropy bound}
\author{Feng-Li Lin \footnote{email address: linfl@phya.snu.ac.kr}}
\address{School of Physics \& Center for Theoretical Physics,
\\ Seoul National University, Seoul 151-742, Korea}
\date{\today}
\maketitle
\begin{abstract}
In this note we calculate the Casimir effect of free thermal
gravitons in Einstein universe and discuss how it changes the
entropy bound condition proposed recently by Verlinde as a higher
dimensional generalization of Cardy's formula for conformal field
theories (CFT).  We find that the graviton's Casimir effect is
necessary in order not to violate Verlinde's bound for weakly
coupled CFT. We also comment on the implication of this new
Cardy's formula to the thermodynamics  of black $p$-brane.
\end{abstract}

\newpage

\section{Introduction}
Verlinde in his recent paper \cite{verlinde} proposed that the
entropy of a D-dimensional conformal field theory (CFT) is given
by the generalized Cardy's formula
\be
S={2\pi R \over D-1}\sqrt{E_c(2E-E_c)}\;, \label{cardy}
\eq
where $R$ is linear size of the system and
\be
E_c \equiv DE-(D-1)TS\;. \label{casimir}
\eq
is the Casimir energy which corresponds to the sub-leading term
in the high T (temperature) expansion of the total energy $E$.

For a given total energy $E$ this formula automatically leads to
the Bekenstein's entropy bound \cite{bekenstein} of the
macroscopic system with limited gravity,
\be
S \le S_B \equiv {2\pi \over D-1}ER \label{bekenstein}\;,
\eq
the bound is saturated when $E_c=E$.

Verlinde has shown that Cardy's formula is exact for the strongly
coupled CFTs by using their holographic dual description
\cite{verlinde,witten}. Moreover, he showed that Cardy's formula
of Eq.(\ref{cardy}) holds even for strongly gravitational system
such as the early universe with the help of a newly proposed
cosmological principle which states that {\it the Casimir energy
itself is not sufficient to form a universe-size black hole}.
Furthermore, the Cardy's formula coincides exactly with the
Friedman equation when the above energy bound is saturated; the
resultant entropy which is called Hubble bound
\cite{veneziano,verlinde} obeys the area law as expected from the
holographic nature of gravity theory \cite{holography,cosmic}.

The Cardy's formula is also checked for weakly coupled CFTs in
\cite{kutasov} by Kutasov and Larsen. They find that the formula
is in general not exact, which results in a violation of
Bekenstein bound in the low energy density.

As shown in \cite{kutasov} the partition function of a free D=4
CFT in Einstein universe can be decomposed into the product of the
basic partition functions as the following:
\be
Z^{(4)}_{CFT}=[Z_b^{(4)}]^A [Z_b^{(2)}]^B \label{cft}
\eq
where
\bee
A&=&n_S+2n_V+{7\over 4}n_F \;,
\\
B&=&-(2n_V+{1\over 4} n_F)\;,
\eqq
for a theory with $n_S$ scalars, $n_F$ Weyl fermions and $n_V$
Maxwell fields. For example, $A=15N$, $B=-3N$ for ${\cal N}=4$
$U(N)$ super-Yang-Mills (SYM) theory.

The basic partition functions are defined by
\be
Z_b^{(d)}=\prod_{n=0}^{\infty}({1 \over 1-q^{n+1}})^{(n+1)^{d-2}}
\label{basic}
\eq
where $q=e^{-2\pi \delta}$ and $\delta=1/(2\pi RT)$ with R the
radius of $S^3$ and $T$ the temperature. For references the
explicit expression of $lnZ_b^{(2,4)}$ in the high T expansion are
\bee
lnZ^{(2)}_b&=&2\pi{1\over 24}(\delta^{-1}-\delta)+{1\over 2}+
ln\delta +\it{O}(e^{-2\pi \delta})\;, \label{z2}
\\
lnZ^{(4)}_b&=&2\pi{1\over 240}({1\over3}
\delta^{-3}+\delta)+\it{O}(e^{-2\pi \delta})\;. \label{sum}
\eqq
The absence of the higher polynomial terms in $\delta$ is due to
the modular invariance of CFT on $S^1 \times S^3$ \cite{cardy1},
that is
\be
I_b^{(d)}({1\over \delta})=(-1)^{{d \over 2}}I_b^{(d)}(\delta)\;,
\eq
where
\be
I_b^{(d)}(\delta)\equiv -\delta^{{d \over 2}}{\partial \over
\partial \delta}lnZ_b^{(d)} \;.
\eq
As will be shown there is no modular invariance for graviton
partition sum.

From Eqs. (\ref{cft}) and (\ref{sum}), we can derive the free
energy $F=-TlnZ^{(4)}_{CFT}$, and the result is
\be
-FR={A\over 720} \delta^{-4}+{B \over 24} \delta^{-2}+ ({A \over
240}-{B \over 24})+ \it{O}(e^{-2\pi \delta}) \;.\label{free}
\eq
We note that the leading term ($\sim \delta^{-4}$ for small
$\delta$) is coming from $lnZ^{(4)}_{b}$ and the sub-leading term
 ($\sim \delta^{-2}$) from $lnZ^{(2)}_b$ which is the leading
Casimir effect. This result is exactly the same as the more
familiar one \cite{myers} derived from path integral using
zeta-function regularization for CFT on general curved background
${\cal M}$
\be
{F\over V}=-{\pi^2 T^4 \over 90}A a_0({\cal M}) -{T^2 \over 24} B
a_1({\cal M})+\cdots\;,
\eq
where $a_k({\cal M})$ are the well-known "Hamidew" coefficients
\cite{myers}. For ${\cal M}=S^1 \times S^3$, $a_0=1, a_1={2\over
R^2}$.

One can then deduce $E=F+TS$ and $S$ from $F$ in the way for a
canonical ensemble and $E_c$ from Eq.
(\ref{casimir})\footnote{For completeness, the explicit
expressions are $S=2\pi [{A \over 180} \delta^{-3}+{B \over 12}
\delta^{-1}]+\it{O}(e^{-2\pi \delta})$, and $E_cR={-B \over
12}\delta^{-2}+({-A \over 60}+{B\over 6})+\it{O}(e^{-2\pi
\delta})$. Note that, the leading term in $E_c$ is zero for
conformal scalars but positive for fermions and gauge fields, and
also for supergravitons as shown later; however, the
$\delta$-independent piece is negative in general, which is the
usual Casimir energy at zero temperature.}. It is easy to see
\cite{kutasov} that Cardy's formula of Eq. (\ref{cardy}) is not
exact; and for the entropy to be bounded by the formula requires
\be
{A \over -B} \le {5 \over 2} \label{bound}
\eq
where the equality holds when the bound is saturated. For ${\cal
N}=4$, $U(N)$ SYM, ${A \over -B}=5$ for all $N$ and thus the bound
is violated. In general we could arbitrarily adjust the matter
content to satisfy the above entropy bound condition, but in this
paper we will consider only ${\cal N}=4$ SYM and see how
graviton's Casimir effect changes the entropy bound condition for
SYM.

Moreover, the authors of \cite{kutasov} observe that if Eq.
(\ref{bound}) does not hold, then the Bekenstein bound of Eq.
(\ref{bekenstein}) will be violated when $ER \lesssim {A \over 9
\times 720}$;  however this condition can be translated into
$\delta \ge (3)^{3/4}$ by using the explicit $\delta$-dependence
of $ER={A \over 240} \delta^{-4}+{\it O}(\delta^{-2})$, which
implies the high $T$ (small $\delta$) expansion of free energy in
Eq. (\ref{free}) is no longer valid. It deserves more study of
the low temperature thermodynamics on the Bekenstein bound.

On the other hand, in the high $T$ regime where Eq. (\ref{free})
works and the Bekenstein bound is not violated, the curvature
effect becomes important because $\delta={1\over 2\pi RT}\ll 1$,
the thermal energy becomes larger than the characteristic
planckian energy which is inversely proportional to $R$. It is
then natural to incorporate the contribution of thermal gravitons
and gravitinos to the total partition function $Z^{(4)}=Z^{(4)}_G
Z^{(4)}_{CFT}$ where $Z^{(4)}_G$ is the partition function due to
gravitons and gravitinos.

\section{Casimir effect of Graviton}

In the following we will calculate $Z^{(4)}_G$ and discuss how it
changes the condition on the entropy bound.  The usual way to
calculate the partition function or the effective action of a
field theory on a fixed background is by evaluating the path
integral up to one-loop \cite{myers}. However, in \cite{kutasov} a
more efficient way for the CFT on $S^1 \times S^3$ is to classify
the operator content by the representations of $SO(4)\simeq SU(2)
\times SU(2)$, the isometry group of $S^3$, and to calculate the
partition sum from it. For example, a conformal scalar and its
higher descendants are represented by $({n\over 2},{n\over2})$ of
$SO(4)$ with degeneracy $(n+1)^2$ and conformal weight
$\Delta=n+1$ for $n=1,2,3,...$, and the resulting partition sum is
\be
Z^{(4)}_S=\prod_{n=0}^{\infty}({1 \over
1-q^{n+1}})^{(n+1)^2}=Z^{(4)}_b\;.
\eq
This method of enumerating the operator content has the advantage
of automatically taking care of the constraints such as equations
of motion, Bianchi identities and etc.

Similarly, the Maxwell field and its descendants are represented
by $({n\over 2},{n+2\over2})+h.c.$ with degeneracy $2(n+1)(n+3)$
and conformal weight $\Delta=n+2$ \footnote{The primary operator is
not $A_{\mu}$ of scaling dimension one but the field strength
$F_{\mu \nu}$ of scaling dimension two because the first is not gauge
invariant but the latter is.}, and the resulting partition sum is
\be
Z^{(4)}_V=\prod_{n=0}^{\infty}({1 \over 1-q^{n+1}})^{2n(n+2)}=
[Z^{(4)}_b]^2[Z^{(2)}_b]^{-2}\;.
\eq
Note that the leading term is just twice the one for the scalar
as expected for massless photon; however, this is not the case
for the leading Casimir effect.

Generalizing the above counting to graviton, the contribution to
the partition sum is due to the spin two representations
$({n\over 2},{n+4 \over 2})+h.c.$ with degeneracy $2(n+1)(n+5)$
and conformal weight $\Delta=n+3$. The scaling dimension of
$\delta g_{\mu \nu}=g_{\mu \nu}-g^{(B)}_{\mu \nu}$ is one, and
from the requirement of general covariance and conformal
invariance the lowest operator should be the Weyl tensor ($\sim
\partial \partial \delta g$) which has ten independent components
\cite{weinberg} and scaling dimension three, this agrees with the
above counting for $n=0$.

The resulting partition sum for graviton is
\be
Z^{(4)}_{g}= \prod_{n=0}^{\infty}({1 \over 1-q^{n+2}})^{2n(n+4)}\;,
\eq
which cannot be decomposed into the basic partition functions of
Eq.(\ref{basic}). Instead we should evaluate the following new
basic partition functions
\be
Z_b^{'(d)}=\prod_{n=0}^{\infty}({1 \over
1-q^{n+2}})^{(n+2)^{d-2}}\;.
\eq

We generalize the method in \cite{cardy1,kutasov} to calculate
$Z_b^{'(d)}$ by the following expansion
\be
-{\partial lnZ_b^{'(d)}\over \partial \delta}=2\pi
\sum_{n=0}^{\infty} (n+2)^{d-1} \sum_{k=1}^{\infty} e^{-2\pi
\delta (n+2) k}\;,
\eq
and using the Mellin representation
\be
e^{-x}={1\over 2\pi i}\int_{C}x^{-z}\Gamma(z)dz
\eq
where the contour $C$ is along the imaginary axis with $Re(z)>0$
large, we arrive
\be
-{\partial \over \partial \delta} lnZ_b^{'(d)}= {1\over 2\pi i}
\int_C (2\pi)^{1-z} \delta^{-z} \zeta(z+1-d) \zeta(z) \Gamma(z) dz
-{1\over 2\pi i} \int_C (2\pi)^{1-z} \delta^{-z} \zeta(z)
\Gamma(z) dz \;.
\eq
It is easy to see that the first term is just the same as
$-{\partial \over \partial \delta} lnZ_b^{(d)}$, and the
integrand of the second term has the poles at
$z=1,0,-1,-3,\cdots$. The resulting expressions of $Z^{'(d)}_b$
in the expansion of $\delta$ are
\bee
lnZ^{'(2)}_b&=&lnZ^{(2)}_b+ln\delta+{\it O}(\delta)=2\pi{1 \over
24} \delta^{-1} + {3\over2} ln\delta + {\it O}(\delta)\;,
\\
lnZ^{'(4)}_b&=&lnZ^{(4)}_b+ln\delta+{\it O}(\delta)=2\pi{1 \over
720} \delta^{-3} + {1\over2} ln\delta + {\it O}(\delta)\;.
\eqq
We see that the high order terms exist because there is no
modular invariance property for the new partition sums; however,
the leading terms here are still the same as those in $Z^{(d)}_b$.
Note that the leading terms are the only relevant terms in
determining the entropy bound condition.

The graviton partition sum $Z^{(4)}_g$ can be decomposed into
\be
Z^{(4)}_g=[Z^{'(4)}_b]^2[Z^{'(2)}_b]^{-8}\simeq
[Z^{(4)}_b]^2[Z^{2}_b]^{-8}\;,
\eq
where "$\simeq$" means having the same leading and sub-leading
terms. Note that the leading term is just twice of the one for
scalar as expected. The resulting $Z^{(4)}_g$ also implies that
the theory consisting of only free conformal thermal graviton
will not violate the entropy bound given by the Cardy's formula
of Eq. (\ref{cardy}) because it has $({A \over -B})_{g}={1\over
4}<{5 \over 2}$.

Similarly, we can calculate the gravitino's partition sum by
identifying its descendants as described by the representation
$2(n,{n+3\over2})$ of $SO(4)$ with degeneracy $2(n+1)(n+4)$ and
conformal weight $\Delta=n+{5\over 2}$. The resulting partition
sum for gravitino is
\be
Z^{(4)}_{gf}=\prod_{n=0}^{\infty}(1+q^{n+{3\over
2}})^{2n(n+3)}=[Z^{'(4)}_f]^2[Z^{'(2)}_f]^{-9/2}\;,
\eq
where the basic fermionic partition functions are defined as
follows:
\be
Z^{'(d)}_{f}=\prod_{n=0}^{\infty}(1+q^{n+{3\over
2}})^{(n+{3\over2})^{d-2}}\;.
\eq

Using the identity
\be
\sum_{n=0}^{\infty} (n+{3\over2})^{-z}=(2^z-1)\zeta(z)-2^z
\eq
and the Mellin representation, it can be shown that the leading
term in $Z^{'(d)}_f$ is the same as in the basic partition
function for a Weyl fermion
\be
Z^{(d)}_f=\prod_{n=0}^{\infty}(1+q^{n+{1\over
2}})^{(n+{1\over2})^{d-2}}=e^{(1-{1\over 2^{d-1}})} Z^{(d)}_b\;.
\eq
We then arrive
\be
Z^{(4)}_{gf}\simeq [Z^{(4)}_b]^{7\over 4}[Z^{2}_b]^{-9 \over 4}\;,
\eq
note that the leading term is the same as the one for a Weyl
fermion.

Combining the contributions of graviton and gravitino together we
find that the total partition function of the on-shell
supergravity theory is
\be
Z^{(4)}_G=Z^{(4)}_g[Z^{(4)}_{gf}]^{\cal N}\simeq
[Z^{(4)}_b]^{2+{7\over4}{\cal N}} [Z^{(2)}_b]^{-8-{9\over 4}{\cal
N}}\;,
\eq
where ${\cal N}$ is the number of supersymmetries. We see that the
entropy bound condition of Eq. (\ref{bound}) is not violated
because $({A\over -B})_{sugra}={2+ {7\over4} {\cal N} \over 8+{9
\over 4}{\cal N} }\le{5\over 2}$.

Now we could combine the contribution of ${\cal N}=4$ $SU(N)$ SYM
and the thermal supergraviton together, it yields
\be
Z^{(4)}=Z^{(4)}_{CFT}Z^{(4)}_G=[Z^{(4)}_b]^{2+{7\over4}{\cal
N}+15N} [Z^{(2)}_b]^{-8-{9\over 4}{\cal N}-3N}\;;
\eq
we see that the entropy bound condition becomes
\be
{A\over -B}={9+15N \over 17+ 3N} \le {5\over 2}
\eq
which leads to a constraint on the rank of the gauge group
\be
N \le {67 \over 15}\;.
\eq
By generalizing to more physical matter contents such as the
Standard model,  one may find a deep connection between entropy
bound and why there are only three colors in nature.

\section{Black p-brane and Cardy's formula}
The $D=2$ Cardy's formula of Eq. (\ref{cardy}) is  the same as the
one derived from the saddle point approximation of the formula for
density of states \cite{cardy2,carlip}
\be
\rho(\Delta)=\int d\delta e^{2\pi\delta \Delta}
[Z^{(2)}_b(\delta)]^{c} \simeq e^{2\pi \sqrt{{c\over
6}(\Delta-{c\over 24})}}
\eq
where c is the central charge; $Z^{(2)}_b$ is given by Eq.
(\ref{z2}) in which the higher order terms is suppressed by the
modular invariance property and thus the saddle point
approximation can be simplified. The entropy $S=ln\rho(\Delta)$
agrees with the Cardy formula of Eq. (\ref{cardy}) by identifying
$E_cR={c \over 12}$ and $ER=\Delta$. For $D > 2$, there is no such
simple square-root Cardy's formula.

On the other hand, the square-root behavior of $D=2$ Cardy's
formula has been shown to be obeyed by the near-horizon classical
gravitational dynamics for the $D>2$ systems such as the black
holes \cite{vafa,strominger0,carlip1,solodukhin}, de Sitter
universe \cite{strominger,lin} and the apparent horizons
\cite{brustein}. These cases implies that the near-horizon physics
is associated with a $D=2$ CFT, and exemplify the holographic
nature of strong gravity regime. It is then curious to see if the
Cardy's formula of Eq. (\ref{cardy}) can be an indication of
holographic nature of strong gravitational system. It is easy to
check that the Cardy's formula is satisfied trivially by the
Schwarzschild black hole but not by the Reissner-Nordstrom black
hole with $E_c\equiv E-TS-\mu Q$ where $\mu$ and $Q$ are the
chemical potential and the corresponding charge. As shown below,
a nontrivial example which fits the Cardy's formula is the black
$p$-brane ($p<7$) described by
\bee
ds^2_{10}&=&(H_p(r))^{-1\over
2}(-f(r)dt^2+d^2x_{\parallel})+(H_p(r))^{1\over 2}({dr^2 \over
f(r)}+r^2 d\Omega^2_{8-p})\;,
\\
C_{012 \cdots p}(r)&=&\sqrt{1+{r^{7-p}_0 \over L^{7-p}}}{H_p(r)-1
\over H_p(r)}\;,
\\
H_p(r)&=&1+{L^{7-p} \over r^{7-p}}\;, \qquad \qquad
f(r)=1-{r_0^{7-p} \over r^{7-p}}\;,
\eqq
where the parameters $L$ and $r_0$ are the anti-de Sitter throat
size and the position of the horizon, respectively. They play
analogous roles of the size and temperature as those in CFT.

From the metric and the $RR$ potential $C_{p+1}$, we can derive
the mass $M$, $RR$ charge $Q$, chemical potential $\mu$,
temperature $T$, and entropy $S$ for the system\cite{kiritsis}:
\bee
M&=&{\Omega_{8-p} V_p \over
2\kappa^2_{10}}[(8-p)r_0^{7-p}+(7-p)L^{7-p}]\;,
\\
Q&=&{(7-p)\Omega_{8-p} \over 2\kappa^2_{10}T^{p}}L^{7-p \over 2}
\sqrt{r_0^{7-p}+L^{7-p}}\;,
\\
\mu&=&V_p T_p {L^{7-p \over 2} \over \sqrt{r_0^{7-p}+L^{7-p}}}\;,
\qquad \qquad T={7-p \over 4\pi} {r_0^{5-p \over 2} \over
\sqrt{r_0^{7-p}+L^{7-p}}}\;,
\\
S&=&{4\pi \Omega_{8-p} V_p \over 2\kappa^2_{10}} r_0^{9-p \over 2}
\sqrt{r_0^{7-p}+L^{7-p}}\;,
\eqq
where $V_p$ is the spatial volume of the $p$-brane, and $T_p$ is
the brane tension. These thermodynamical quantities satisfy the
first law: $dM= TdS+\mu dQ$; however in its integral form there
is an excess which can be identified as the Casimir energy in
\cite{verlinde}
\be
E_c=2(M-TS-\mu Q)\;.
\eq

It is then straightforward to see that
\be
S={2\pi r_0 \over \sqrt{7-p}}\sqrt{E_c(2M-E_c)}\;, \label{sp}
\eq
which is in the same form of Cardy's formula of Eq. (\ref{cardy})
but with different overall factors. It is interesting to see if
this formula bears any microscopic interpretation from D-brane and
string theory as done for black string \cite{horowitz}. It also
deserves further study of the thermodynamics of the holographic
dual theory corresponding to the black $p$-brane and examine the
validity of Eq. (\ref{sp}) from the dual point of view.


\section{Conclusion}
In Verlinde's proposal \cite{verlinde} the Cardy's formula of Eq.
(\ref{cardy}) is expected to be exact for CFT, but it turns out
that this is true only for strongly coupled theory but not for
weakly coupled one \cite{kutasov}. On the other hand, this
formula unifies the Bekenstein bound and Hubble bound in the
cosmology context. As long as the theory does not violate the
entropy bound given by Cardy's formula, it would have no problem
to satisfy both Bekenstein bound for weakly gravity regime  and
Hubble bound for strong gravity regime along the evolution of the
closed universe. From these facts it is more natural to think
Verlinde's proposal as an universal entropy bound but not an
exact entropy formula for CFT. We have seen that the pure
perturbative effect of gravity will not violate the bound by
Cardy's formula. This can be viewed as a self-consistent test for
the Bekenstein entropy bound though that its validity has been
under debate \cite{debate} since it was proposed twenty years ago.
Moreover, when combining with CFT's contribution, we see that
graviton's Casimir effect is necessary for the CFT to satisfy the
entropy bound condition which yields a constraint on the rank of
the gauge group of the CFT. We also see  that there is an
intriguing resemblance of Cardy's formula in black $p$-brane's
thermodynamics, which deserves further study for its  physical
implication.

\bigskip

{\bf Acknowledgement}: The author would like to thank Pei-Ming Ho,
Choonkyu Lee, Miao Li and Soo-Jong Rey for helpful discussions,
also for the warm hospitality of NTU string theory group
where this work is finalized. This work is supported by BK-21
Initiative in Physics (SNU - Project 2), KRF International
Collaboration Grant, and KOSEF Interdisciplinary Research Grant
98-07-02-07-01-5.

\end{document}